\documentclass[aps,prl,preprint,showkeys,groupedaddress]{revtex4-1}
\usepackage[usenames,dvipsnames]{color}
\usepackage{rotating,ctable,capt-of,xcolor}

\definecolor{darkyellow2}{RGB}{153,153,0}
\definecolor{darkyellow}{RGB}{255,220,0}
\definecolor{darkorange}{RGB}{235,102,0}
\definecolor{darkgreen}{rgb}{0.05, 0.5, 0.06}
\definecolor{my_lightgray}{gray}{0.3}
\definecolor{darkmagenta}{rgb}{0.85, 0.2, 0.53}
\definecolor{matlab1}{rgb}{0,0.4470,0.7410}
\definecolor{matlab2}{rgb}{0.8500,0.3250,0.0980}
\definecolor{matlab3}{rgb}{0.9290,0.6940,0.1250}
\definecolor{matlab4}{rgb}{0.4940,0.1840,0.5560}
\definecolor{matlab5}{rgb}{0.4660,0.6740,0.1880}
\definecolor{matlab6}{rgb}{0.3010,0.7450,0.9330}
\definecolor{matlab7}{rgb}{0.6350,0.0780,0.1840}

\newcolumntype{L}[1]{>{\raggedright\let\newline\\\arraybackslash\hspace{0pt}}m{#1}}
\newcolumntype{C}[1]{>{\centering\let\newline\\\arraybackslash\hspace{0pt}}m{#1}}
\newcolumntype{R}[1]{>{\raggedleft\let\newline\\\arraybackslash\hspace{0pt}}m{#1}}

\begin{document}
%\title{Turbulence transition in a sudden expansion pipe} %for header on odd pages
%\shortauthor{M. Q. Nguyen, B. Lebon, M. S. Shadloo, J. Peixinho and A. Hadjadj} 
%for header on even pages

\title{Features of Transition to Turbulence in Sudden Expansion Pipe Flows}
\author{ Minh Quan Nguyen}
\affiliation{UMR 6614-CORIA, Normandie University, CNRS-University and INSA of Rouen, 76000 Rouen, France}

\author{ Benoit Lebon}
\affiliation{Laboratoire Ondes et Milieux Complexes, CNRS and Universit\'e Le Havre Normandie, 76600 Le Havre, France}

\author{ Mostafa Safdari Shadloo}
\affiliation{UMR 6614-CORIA, Normandie University, CNRS-University and INSA of Rouen, 76000 Rouen, France}

\author{ Jorge Peixinho}
\affiliation{Laboratoire Ondes et Milieux Complexes, CNRS and Universit\'e Le Havre Normandie, 76600 Le Havre, France}

\author{ Abdellah Hadjadj}
\affiliation{UMR 6614-CORIA, Normandie University, CNRS-University and INSA of Rouen, 76000 Rouen, France}

\begin{abstract}
The complex flow features resulting from the laminar-turbulent transition (LTT) in a sudden expansion pipe flow, \textcolor{black}{with expansion ratio of 1:2} subjected to an inlet vortex perturbation is investigated by means of direct numerical simulations (DNS). \textcolor{black}{It is shown that the threshold for LTT  described by a power law scaling with -3 exponent that links the perturbation intensity to the subcritical transitional Reynolds number.} Additionally, a new type of instability is found within a narrow range of flow parameters. This instability  \textcolor{black}{originates} from the region of intense shear rate which is a result of the flow symmetry breakdown. Unlike the fast transition, usually reported in the literature, the new instability emerges gradually from a laminar state and appears to be chaotic and strongly unsteady. Additionally, \textcolor{black}{the simulations show a hysteresis mode transition due to the reestablishment of the recirculation zone in a certain range of Reynolds numbers.} The latter depends on (i) the initial and final quasi-steady states, (ii) the observation time and (iii) the number of intermediate steps taken when increasing and decreasing the Reynolds number.
\end{abstract}
\pacs{}

\keywords{Transition to turbulence, Sudden expansion pipe flow, Instability control, Direct numerical simulation (DNS)}

\maketitle
\section{Introduction}

The flow through an axisymmetric sudden expansion in a circular pipe is a basic configuration, which occurs in many industrial applications, such as heat exchanger, mixing chamber, combustion chamber, etc.. This basic geometry is also used as a building block to model more complex flows such as those occurring in arterial stenoses  \citep{Pollard1981}, pistons \citep{Boughamoura2003}, and transportation pipes \citep{Koronaki2001}, among others. In these applications, the capacity of predicting when the flow will become turbulent is crucial. In the literature, there are many efforts in theoretical analysis \citep{Teyssandiert1974}, experimental explorations \citep{Back1972,Latornell1986} and numerical simulations \citep{Macagno1967,Varghese2007} focusing on this problem, or a similar geometries such as the planar expansion \citep{Fearn1990,Baloch1995} and the abrupt contraction and expansion \citep{Xia1992,Varghese2007,Bertolotti2001}. 
\textcolor{black}{Sudden expansion with various expansion ratio in investigated by \citet{Lanzerstorfer2012}. More recently, \citet{Lebon2018} found a new mechanism for periodic bursting of the recirculation region in the flow in a circular pipe with the expansion ratio of 1:2. Yet,} a consensus about the sequence of event in the transition from laminar to turbulence seems to be relatively well-established, but the exact value of critical Reynolds number is still not firmly determined, and the different transition scenarios are not yet fully elucidated.

The flow is mainly controlled by the inlet Reynolds number, $Re=\overline{U}d/\nu$, where $\overline{U}$ is the mean velocity at the inlet,  $d$ is the inlet diameter and $\nu$ is the fluid kinematic viscosity. In the laminar state, the flow is axisymmetric. As $Re$ increases, the flow starts to break its symmetrical properties but remains steady until $Re = 1139\pm 10$ as reported in the experiments by Mullin et al. \cite{Mullin2009}. Noted that the onset of symmetry breakdown is at much lower value for the case of channel sudden expansion, as reported to be $Re\approx40$ by Fearn et al. \cite{Fearn1990} and $Re=216$ by Drikakis \cite{Drikakis1997}. In the sudden circular pipe expansion flows, intermittent bursts were \textcolor{black}{reported  \cite{Latornell1986,Mullin2009} to appear at $Re=635$ and $Re\approx 1453\pm 41$.} Then the flow enters to an oscillating state. \textcolor{black}{For instances, this state found at $Re\approx 1567 \pm 16$ \cite{Mullin2009}, $1500<Re<1700$ \cite{Sreenivasan1983}, and $Re=750$ \cite{Latornell1986} among others.} Finally, the oscillating pattern breaks into a chaotic movement at even higher Reynolds numbers.

More recently, a global stability analysis was performed by Sanmiguel-Rojas et al. \cite{rojas2010}. They showed that the flow can remain symmetry up to $Re\approx 3273$, which is much larger than the value found experimentally. Subsequent simulations by  Cliffe et al. \cite{Cliffe2011} also indicated that the steady supercritical bifurcation point lies at even higher Reynolds numbers, $i.e.$ $Re\approx 5000$. These discrepancies among the reported results suggest that there might be some missing underlying parameters. 

A detailed study of transient growth stability was performed by  Cantwell et al. \cite{Cantwell2010}. They illustrated that the flow is linearly stable up to $Re = 1400$, however, even in the subcritical regime, the sudden expansion can amplify the \textcolor{black}{energy of the} perturbation up to 6 order of magnitude in the inlet and then decay. This difference may be explained by the fact that in experimental studies, as warned by \citet{Cantwell2010}, the imperfection of apparatus has a strong impact on the measured critical Reynolds number. Therefore, the values of critical Reynolds number seem to be dependant both on the perturbation nature and its amplitude. In this case, a numerical simulation with a well-defined finite amplitude perturbation is required to better understand the underlying physics. 

The first direct numerical simulation (DNS) of finite amplitude perturbation in sudden expansion flow was recently performed by Sanmiguel-Rojas \& Mullin \cite{rojas2012} where many interesting results are reported. They showed that for a given Reynolds number there exists a region where the flow in laminar state can be forced to enter turbulence state with a finite perturbation. The minimal value of perturbation required to trigger the turbulence, scales accurately with $Re^{-0.006}$. This result has an important role in the passive flow control, however, it arises many questions. For instances, is this result universal? Will another kind of perturbation have the same behaviour? Is the scaling law still valid? The question is crucial especially by knowing that the previous perturbation scheme used by Sanmiguel-Rojas \& Mullin \cite{rojas2012} is simply a tilt disturbance in the velocity field. Therefore, it does not respect the no-slip boundary condition at the inlet wall. Another interesting result is: when they \textcolor{black}{increased and then decreased} the Reynolds number, a hysteresis behaviour was found for the Reynolds numbers ranging from $1450$ to $1850$. This result has strong implications on the nature of transition. This important point deserves more attention, since in the original work, the process of variation of Reynolds number as well as the physical time of the reported flow state are not specified. One may ask, will the hysteresis not occur if the Reynolds number vary in quasi-static manner? Or will the results change if the observation time is different? \textcolor{black}{In the present study, the authors propose a similar study, but with vortex disturbance, to revisit the power law and the hysteresis behaviors. }

\section{Numerical set-up}
The present work focuses on a pipe flow with a sudden expansion with the expansion ratio of 1:2. The fluid flow system is solved using Nek5000 \citep{nek5000}, a well-validated high-order spectral element code for transitional and fully turbulent flows \citep{Ducoin2017,Selvam2015}. The governing equations are mass and momentum conservations in an isothermal incompressible limit:

\begin{equation}
\nabla \cdot \mathbf{u}=0
\end{equation}

\begin{equation}
\frac{\partial}{\partial t}\mathbf{u}+\mathbf{u}\cdot\nabla \mathbf{u}=-\nabla p+\nu\Delta \mathbf{u}
\end{equation}
Where $\mathbf{u}$ is the velocity field, $p$ is the pressure, and $t$ is the physical time. \textcolor{black}{The density of flow is set to unity for simplicity.}

\begin{figure}
\centering
\includegraphics[scale=1]{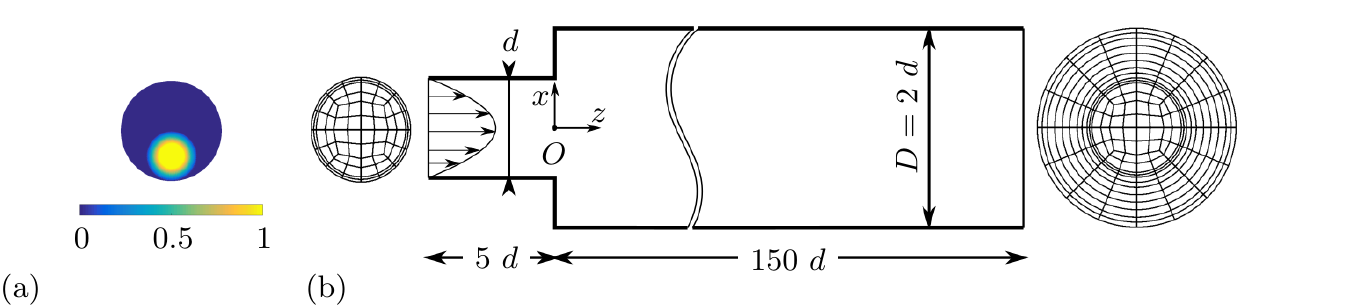}
\caption{(a) Axial vorticity of the vortex perturbation ($R_{\Omega}=d/4$, $(x_{\Omega},y_{\Omega})=(0,-d/4)$ and $\mathcal{A}=1$) and (b) Sketch of fluid domain with 2 cross-sections inlet and outlet mesh. 
\label{Geom}}
\end{figure}

For the expansion pipe test case, the computational domain is axisymmetric (see figure \ref{Geom}). The region upstream of the expansion is called inlet and has the diameter of $d$ and the length of $5d$ . The downstream is the region where we are focusing our analyse on. As illustrated in figure \ref{Geom}, this region has the diameter of $D=2d$  and the length of $L=150d$. The whole domain contains $62~300$ spectral elements where each element consists of $P^3$ Gausse-Legendre-Lobatto (GLL) points, $P$ being the polynomial order.  The element with $P=5$, same as \cite{Selvam2016,Lebon2018}, with total number of 7.9 million calculating points is used in the whole domain for the reported results. However, to confirm our finding, additional simulations are performed with $P=6$ (13.5 million GLL points). A classical set up for the inlet velocity, located at $z=-5d$, is the Hagen-Poiseuille profile, which should satisfy the non slip condition at inlet wall: $\mathbf{u}_{inlet}^{o}(r=d/2,z=-5d)=0$ when $r=\sqrt{x^2+y^2}=d/2$ : 
%\begin{equation}
%\mathbf{u}_{inlet}(x,y)=\left( \begin{array}{ccc}
%0 \\
%0 \\
%(2-1/2*\sqrt{x^2+y^2}) \end{array} \right),
%\end{equation}
\begin{equation}
\mathbf{u}_{inlet}^{o}(x,y,z=-5d)=2[1-4(x^2+y^2)/d^2]  \overline{U} \mathbf{e_z},
\end{equation}
where $(\mathbf{e_x},\mathbf{e_y},\mathbf{e_z})$ are the set of 3 unit vectors of Cartesian base, respectively. The length unit  $d$ and the time unit $s$ are chosen such that the mean inlet velocity $\overline{U}$ reproduces unity. 

In order to trigger turbulence in subcritical Reynolds numbers, a vortex perturbation is added to the inlet parabolic profile. This modifies the expression of velocity inlet as:
\begin{equation}
\label{inlet_profile}
\mathbf{u}_{inlet}(x,y,z=-5d)=\mathbf{u}_{inlet}^{o}+\mathcal{A}\Omega\left( \begin{array}{ccc}
- y \\
 x\\
0 \end{array} \right)=\left( \begin{array}{ccc}
-\mathcal{A} \Omega y \\
\mathcal{A} \Omega x\\
2(1-4(\underbrace{x^2+y^2}_{r^2})) \end{array} \right),
\end{equation}
%\begin{equation}
%\mathbf{u}_{inlet}(x,y)=\left( \begin{array}{ccc}
%-\delta \Omega y \\
%\delta \Omega x\\
%(2-1/2*\sqrt{x^2+y^2}) \end{array} \right),
%\end{equation}
where $\mathcal{A}$ and $\Omega$ are the amplitude and the intensity of the vortex perturbation, respectively. The value of $\Omega$ is controlled by its radius, $R_{\Omega}=d/4$, and its center $(x_{\Omega},y_{\Omega})$. Let $r_{\Omega}$ be the distance from a point $(x,y)$ to the center of the vortex: $r_{\Omega}=\sqrt{(x-x_{\Omega})^2+(y-y_{\Omega})^2}$, then the expression of $\Omega(x,y)$ is given by:
\begin{equation}
\Omega(x,y)= \left\{\begin{array}{c c}
1&r_{\Omega} \leq 0.125 \\
2(1-4r_{\Omega})&0.25\geq r_{\Omega}>0.125 \\
0&r_{\Omega}>0.25 \end{array} \right. .
\end{equation}
The parameter of $\Omega(x,y)$  is fixed with $x_{\Omega}=-0.25d$ and $y_{\Omega}=0$, 
\textcolor{black}{ such as the inlet profile respect the incompressible, non slip and non penetrate boundary condition. Indeed, the incompressibility is verified by: 
\begin{equation}
\nabla\cdot\textbf{u}_{inlet}= \left\{\begin{array}{c c}
-\frac{\partial(\mathcal{A} y)}{\partial x}+\frac{\partial(\mathcal{A} x)}{\partial y} +  \frac{2\partial (1-4(x^2+y^2))}{\partial z} = 0 & r_{\Omega} \leq 0.125 \\
2\frac{\partial(\mathcal{A} yr_{\Omega})}{\partial x}-2\frac{\partial(\mathcal{A} xr_{\Omega})}{\partial y} +  \frac{2\partial (1-4(x^2+y^2))}{\partial z} =0 & 0.25\geq r_{\Omega}>0.125 \\
\frac{2\partial (1-4(x^2+y^2))}{\partial z} = 0&r_{\Omega}>0.25 \end{array} \right. . 
\end{equation}
The radius and the position of the vortex is chosen such that $\Omega$ vanishes at the boundary. For the closest point of the boundary to the center of the vortex: $x=-0.5d$ and $y=0$, we have $r_{\Omega}=0.25d$, thus $\Omega(x,y)=0$.} By injecting $\Omega=0$ into equation \ref{inlet_profile}, one can easily verify that $\mathbf{u}_{inlet}$ vanishes at the boundary as well.

%\textcolor{blue}{The parameters are chosen so the vortex disturbance remains in the flow domain, is easily observable and efficiently initiate turbulence.}

 Preliminary tests with the vortex disturbance on the pipe centreline indicate the required amplitude to initiate turbulence was very large \cite{Wu2015}. It corresponds to the limit case of the rotating Hagen-Poiseuille flow discharging into a sudden expansion \cite{miranda-Barea2015}. By positioning the vortex at $(x_\Omega,y_\Omega)=(-0.25d,0)$, the vortex breaks the axial symmetry of the flow and naturally deflects the recirculation region. 
The distribution of $\Omega(x,y)$ can be visualised in figure \ref{Geom}(a). 

\begin{table}
\center
\begin{tabular}{ccccc|ccccc}
Name & $Re$ & $\mathcal{A}$ & Initial condition & Observed states & Name & $Re$ & $\mathcal{A}$ & Initial condition & Observed states  \\[3pt]
1	& 1100 	& 0 		& 	H-P		& LS		 &   8 	& 1600 	& 0 		& 	H-P		& LS		\\
1a 	& 1100 	& 0.458 	& 	1		& LS		  &  8a 	& 1600 	& 0.123	& 	8		& LS		 \\
1b 	& 1100 	& 0.480 	& 	1		& LS, US2	  &  8b 	& 1600 	& 0.128	& 	8		& LS, US2	 \\
1c 	& 1100 	& 0.494 	& 	1		& US1		  &  8c 	& 1600 	& 0.16	& 	8		& US1		 \\
2 	& 1300 	& 0 		& 	H-P		& LS		    & 9a	& 1700	& 0.2	&	9e		& US1 \\
2a 	& 1300 	& 0.2 	& 	2		& LS		 & 9a  	& 2000 	& 0.0773	& 	9		& LS		 \\
2b 	& 1300 	& 0.239	& 	2		& LS, US2	 & 9b  	& 2000 	& 0.0782	& 	9		& LS, US2	\\
2c 	& 1300 	& 0.385	&	2		& US1		& 9c  	& 2000 	& 0.09 	& 	9		& LS, US2	 \\
3a 	& 1325 	& 0.2 	& 	2a		& LS		& 9d  	& 2000	& 0.1		& 	9		& US2		 \\
4a 	& 1350 	& 0.2 	& 	3a		& LS		 & 9e  	& 2000 	& 0.2 	& 	9		& US1		 \\
5a 	& 1360 	& 0.2 	& 	4a		& LS, US2	 & 9f  	& 2000 	& 0.5 	& 	9		& US1		 \\
6a 	& 1375 	& 0.2 	& 	4a		& LS, US2	& 10a  	& 1700 	& 0.2 	& 	3a		& US1		 \\
7a 	& 1400 	& 0.2 	& 	6a		& LS, US2	& L1d  	& 1700 	& 0.2 	& 	9e		& US1 \\
7b 	& 1400 	& 0.2 	& 	2a		& LS, US2	 & L2d  	& 1350 	& 0.2 	& 	L1d		& US2 \\
6b 	& 1375 	& 0.2 	& 	7b		& LS, US2	& L3d  	& 1000 	& 0.2 	& 	L2d		& LS \\
5b 	& 1360 	& 0.2 	& 	6b		& LS, US2	 & L2i  	& 1350 	& 0.2 	& 	L3d		& US2 \\
4b 	& 1350 	& 0.2 	& 	6b		& LS, US2	 & L1i  	& 1700 	& 0.2 	& 	L2i		& US1 \\
3b 	& 1325 	& 0.2 	& 	4b		& LS, US2	 & L3dbis 	& 1325 	& 0.2 	& 	L2d		& LS \\
2b 	& 1300 	& 0.2 	& 	3b		& LS, US2

%%%%%%%%%%%%%%%%%%%

%\hline
%1	& 1100 	& 0 		& 	H-P		& LS		 \\  
%1a 	& 1100 	& 0.458 	& 	1		& LS		  \\  
%1b 	& 1100 	& 0.480 	& 	1		& LS, US2	  \\  
%1c 	& 1100 	& 0.494 	& 	1		& US1		  \\  
%2 	& 1300 	& 0 		& 	H-P		& LS		    \\
%2a 	& 1300 	& 0.2 	& 	2		& LS		\\
%2b 	& 1300 	& 0.239	& 	2		& LS, US2	\\
%2c 	& 1300 	& 0.385	&	2		& US1		\\
%3a 	& 1325 	& 0.2 	& 	2a		& LS		\\
%4a 	& 1350 	& 0.2 	& 	3a		& LS		 \\
%5a 	& 1360 	& 0.2 	& 	4a		& LS, US2	 \\
%6a 	& 1375 	& 0.2 	& 	4a		& LS, US2	\\
%7a 	& 1400 	& 0.2 	& 	6a		& LS, US2	\\
%7b 	& 1400 	& 0.2 	& 	2a		& LS, US2	 \\
%6b 	& 1375 	& 0.2 	& 	7b		& LS, US2	\\
%5b 	& 1360 	& 0.2 	& 	6b		& LS, US2	 \\
%4b 	& 1350 	& 0.2 	& 	6b		& LS, US2	 \\
%3b 	& 1325 	& 0.2 	& 	4b		& LS, US2	 \\

%2b 	& 1300 	& 0.2 	& 	3b		& LS, US2	 \\
%8 	& 1600 	& 0 		& 	H-P		& LS		\\
%8a 	& 1600 	& 0.123	& 	8		& LS		 \\
%8b 	& 1600 	& 0.128	& 	8		& LS, US2	 \\
%8c 	& 1600 	& 0.16	& 	8		& US1		 \\
%9a	& 1700	& 0.2	&	9e		& US1 \\
%9a  	& 2000 	& 0.0773	& 	9		& LS		 \\
%9b  	& 2000 	& 0.0782	& 	9		& LS, US2	\\
%9c  	& 2000 	& 0.09 	& 	9		& LS, US2	 \\
%9d  	& 2000	& 0.1		& 	9		& US2		 \\
%9e  	& 2000 	& 0.2 	& 	9		& US1		 \\
%9f  	& 2000 	& 0.5 	& 	9		& US1		 \\
%10a  	& 1700 	& 0.2 	& 	3a		& US1		 \\
%
%L1d  	& 1700 	& 0.2 	& 	9e		& US1 \\
%L2d  	& 1350 	& 0.2 	& 	L1d		& US2 \\
%L3d  	& 1000 	& 0.2 	& 	L2d		& LS \\
%L2i  	& 1350 	& 0.2 	& 	L3d		& US2 \\
%L1i  	& 1700 	& 0.2 	& 	L2i		& US1 \\
%L3dbis 	& 1325 	& 0.2 	& 	L2d		& LS \\

%\hline
\end{tabular}
\caption{Summary the simulations. Abbreviations: LS - laminar state and US - unsteady state. The initial condition could be either by default Hagen-Poiseuille profile, noted as ``H-P", or the last instance of an another past simulation.}  
\label{Recap_simu} 
\end{table}

\section{Distinction between different states of instabilities}

Laminar or turbulent nature of a flow can be clearly distinguished from the time evolution of the drag coefficient:
   \begin{equation}C_z(t)=\frac{\nu}{4\pi d^2 L \overline{U}^2}\int\limits_{z=0}^{L}\int\limits_{\theta=0}^{2\pi}\left[\frac{\partial{u}_z}{\partial r}\right]_{r=d}r  \: d\theta \: dz,
 \end{equation}
where $d$ and $L$ are, respectively, the radius and the length of the downstream pipe. Here $r=\sqrt{x^2+y^2}$ and $\theta=\rm{arctan}(y/x)$ are the positions in the radial and azimuthal direction in the Cylindrical coordinate system, respectively.

Table \ref{Recap_simu} presents a summary of the selected simulations along with their Reynolds numbers, perturbation intensities, initial conditions and observed states. \textcolor{black}{While the laminar state is noted as LS, there exist 2 types of unsteady states with distinguished amplitude levels and positioning of turbulence puffs}. Therefore, they will be called by a more general name, $i.e.$ unsteady state (US). There are two fundamental differences between LS and US when observed directly from $C_z(t)$. Firstly, LS is steady, therefore, the time variation of the skin-friction coefficient should be negligible, $i.e.$ $dC_z/dt\approx 0$. This is not the case for US. Secondly, LS has a significantly smaller drag coefficient with regards to its US counterpart, $i.e.$ $C_{z,LS}<<C_{z,US}$. 
%One can observe both of these 2 different clearly in figure ... where the time evolutions of $C_z$ of different cases are presented. We show the comparison between 2 case of same $Re$ but start with different initial condition ($Re$ increasing in blue and decreasing in red) 

To better observe the unsteady patterns, the velocity fluctuations are extracted in the pipe centerline by subtracting the instantaneous field $u_z$ from a reference field $u_{zo}$ in the same spatial location, $i.e.$ 
\begin{equation}
u'_z(0,0,z,t)=u_z(0,0,z,t)-u_{zo}(0,0,z),
\end{equation}
and plotted in a space-time diagram. The reference value $u_{zo}$ is the ``closest" available laminar state for a given sets of parameters. In other words, $u_{zo}$ is chosen as the initial condition or the final state of the simulation, if it starts or ends with a laminar state. However, if the simulation starts and ends with turbulence states, a laminar state from the closest $Re$ and/or $\mathcal{A}$ is chosen as the reference. 
The process of constructing such diagram is as the following: at a given instance, the value of interested quantities, here is $u_z$, of $1500$ points in the centreline in $Z$ direction is recorded at each $100$ time-step. The final result is a 2D function of space and time with the resolutions of \textcolor{black}{$0.1d$ and $0.1s$)}, respectively. 
This diagram can provide complementary information to the plot of $C_z$, such as spatial distribution and local intensity of unsteady pattern. 

\begin{figure}
\centering

\includegraphics[scale=1,,trim={0cm 0.6cm 0cm 0cm},clip]{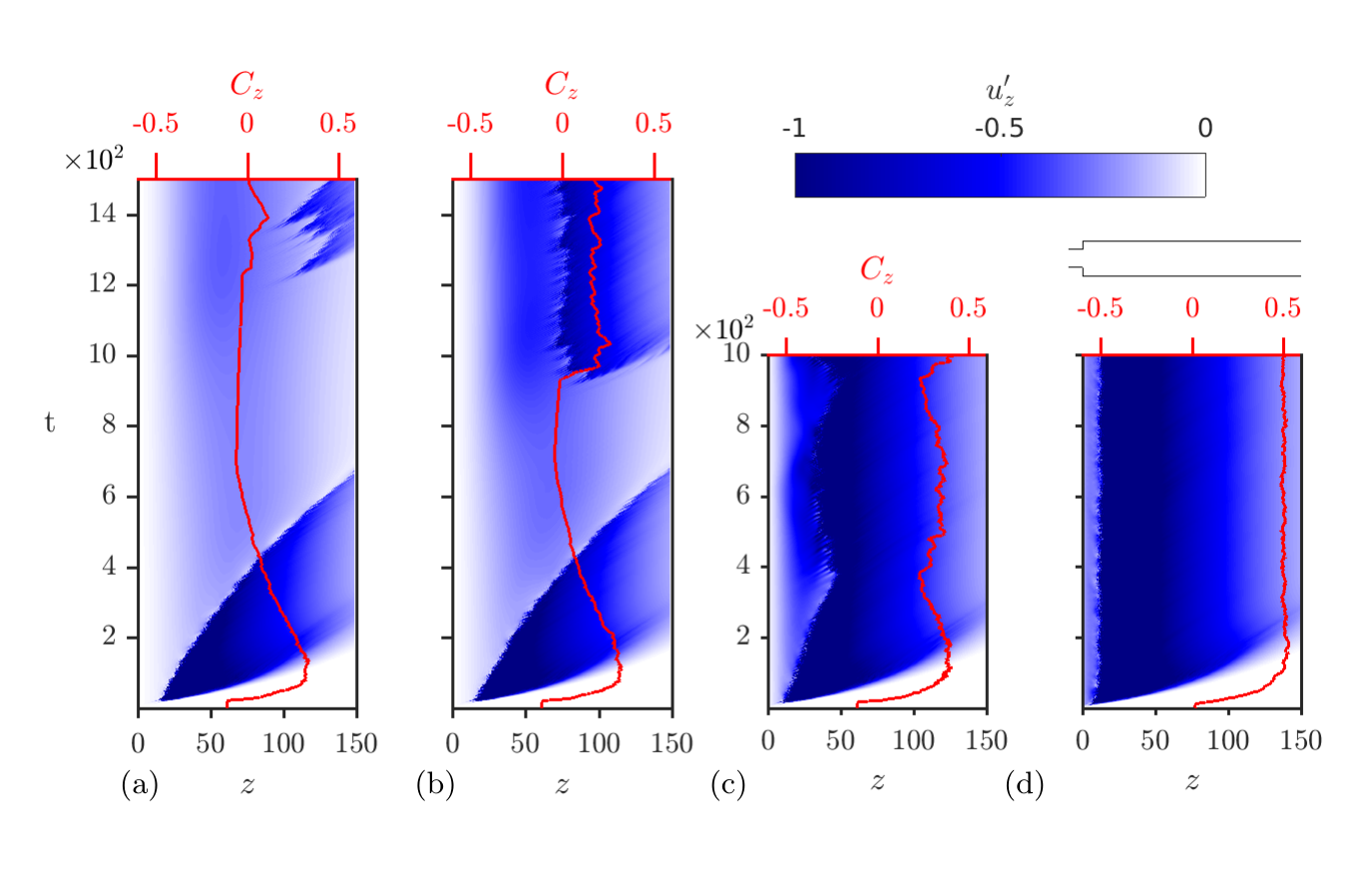}
\caption{Space-time diagrams of the centreline perturbed streamwise velocity (blue colour) of the  \textcolor{black}{$L=150d$ pipe and drag coefficients in red line indicated on the top for $Re = 2000$ at (a) $\mathcal{A}=0.0782$, (b) $\mathcal{A}=0.09$, (c) $\mathcal{A}=0.2$ and (d) $\mathcal{A}=0.5$.} All the diagrams use the colour code on the top left corner, such that light corresponds to laminar flow. \label{Spacetimediagram} }
\end{figure}

Figure \ref{Spacetimediagram} presents the evolution of drag coefficient $C_z$ as a function of time for four different perturbation amplitudes $\mathcal{A}$ at $Re=2000$. The background colour in subfigures shows the space-time diagram. It is noted that the intensity of colour in the diagrams increases with the deceleration of the streamwise velocity. There are two main mechanisms that cause this deceleration. The first one is when the perturbation breaks the flow symmetry down. As a result, the centroid maximum value in the velocity profile is moved away from the centerline. This causes the streamwise velocity in the center of the pipe seemed to be decreased. If the flow is steady, but the symmetry is broken, a smooth region of blue colour will be observed. The second mechanism is when the flow becomes turbulent. As a consequence, the instantaneous streamwise velocity will fluctuate with strong amplitude. For the sake of simplicity, only deceleration is shown in the colormap. Additionally, the mean velocity profile is flattened when the flow becomes turbulent, which also results in a deceleration in the pipe centerline. Hence, the beginning of the turbulence has a clear signature in space-time diagram. This region is found with more intense blue colour, and a noisy interface in space-time diagram. Furthermore, it is obvious from this figure that the emergence of unsteady pattern is closely linked to the rise of drag coefficient $C_z$. Note that at the beginning of each simulation, if there is any change in the parameters \textcolor{black}{comparing to the previous simulation, $i.e.$ the simulation in which the last instance is used as initial condition for the new simulation}, the flow might develop an instability. This instability gets carried downstream and decays. Since the cause of this instability is not physical, it will not be considered here. 

\begin{figure}
\centering
\includegraphics[scale=1]{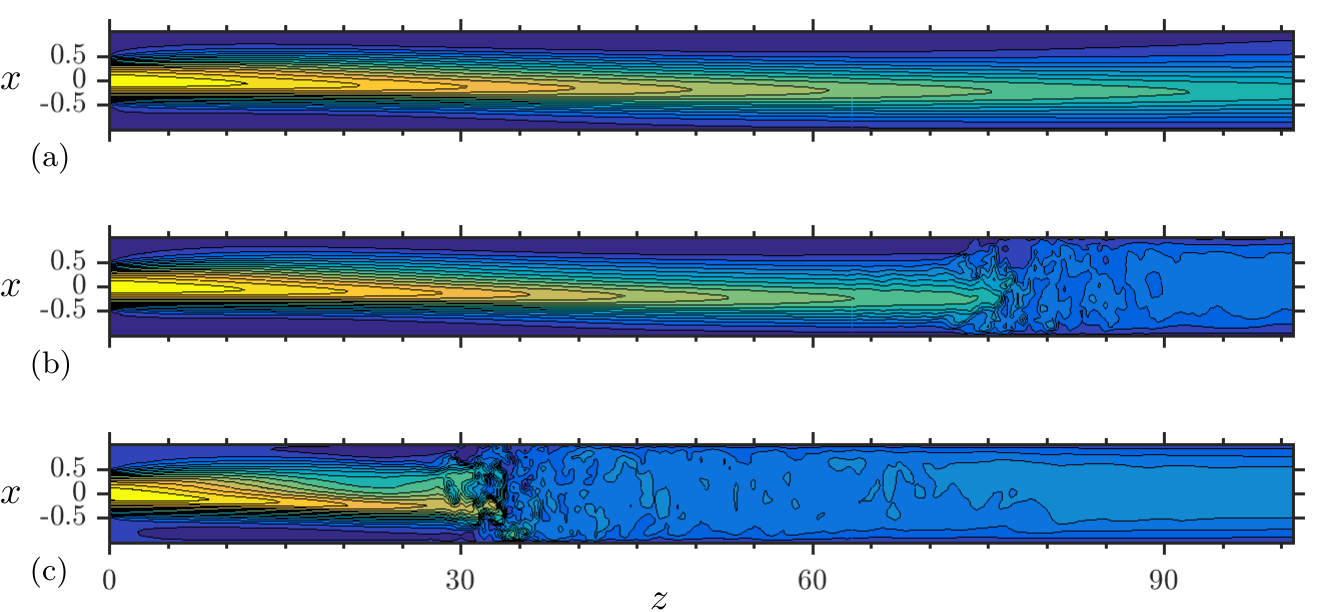}
\caption{Contour plot of instantaneous slide of streamwise velocity, the slide is taken at $y=0$ and zoomed into the range of $0d<z<100d$. Three subplot respectively corresponding to: (a) LS at $Re=2000$, $\mathcal{A}=0.09$, $t=750\ $(s), (b) US2 at $Re=2000$, $\mathcal{A}=0.09$, $t=1250\ $ (s) and (c) US1 at $Re=2000$, $\mathcal{A}=0.2$, $t=1000\ $(s). \label{contourplot}}
\end{figure}

Several unsteady behaviours can be observed in the chosen range of parameters.
\textcolor{black}{For all the cases, the flow exhibit a recirculation zone, which extent depend on $Re$. The recirculation zone is delimited by a reattachement point, the process of determination of this point is simple in laminar case: we find the first point near the wall where the velocity change the sign. For the turbulence cases, an average are over a brief periode of time is needed to determinate the point.}  
For the case corresponding to $Re=2000$ and $\mathcal{A}=0.5$ (see figure \ref{Spacetimediagram}d), the unsteady behaviour starts almost immediately at the beginning of the simulation. The generation and sustention of the instability at the trailing edge seems to be linked with the breaking of the recirculation zone. The flow suddenly loses its smoothness and finds a new reattachment point, with the position right before the trailing edge. At $\mathcal{A}=0.5$, this position is fixed over time. As the perturbation amplitude decreases, the flow has a tendency to restore the recirculation zone, and pushes the trailing edge further downstream. On the other hand, the unsteady pattern has tendency to profilate upstream and downstream, therefore try to push trailing edge upstream. The competition between these two mechanisms might lead to an unsteady position of the trailing edge such as that in the case of $Re=2000$ and $\mathcal{A}=0.2$ (see figure \ref{Spacetimediagram}(c)). The recirculation can be observed clearly with the contour plot of velocity in streamwise direction. 
As shown in figure \ref{contourplot}(c), the unsteady pattern invades and pushes the reattachment point upstream compare to the laminar configuration ($i.e.$ figure \ref{contourplot}(a)). This type of unsteady behaviour will be called the primary unsteady-state pattern, hereafter labelled by US1. Since US1 breaks the recirculation region significantly and gains energy from this process, the transition from LS to US1 is irreversible and a hysteresis behaviour might be expected for this state.
%Since the unsteady pattern can gain energy from the recirculation bubble breaking process, the turbulence fluctuations could become self-sustainable. As consequence the wall shear stress, $\tau_L$, might also become very large. 

\begin{figure}
\centering
\includegraphics[scale=1,trim={0.0cm 0cm 0cm 0cm},clip]{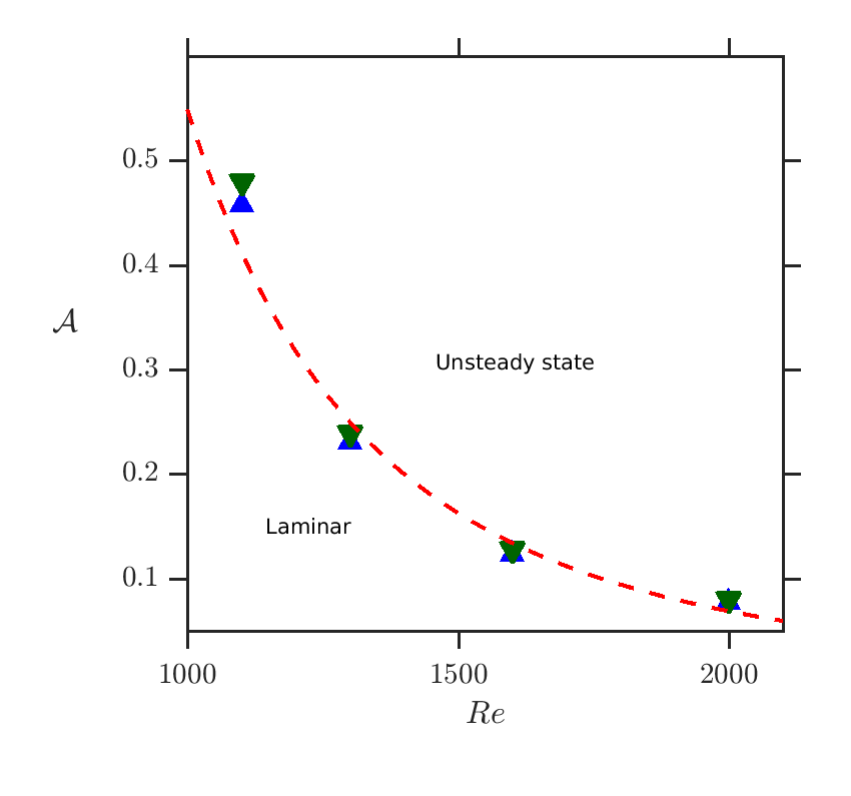}

\caption{Boundary of transition, with green triangle down: US, blue triangle up: LS. Red dashed line is power law $\mathcal{A}\sim Re^{-3}$. \label{Transition_boundary} }   

\end{figure}

For lower values of $\mathcal{A}$, $i.e.$ $\mathcal{A} \leq 0.09$ (see figures \ref{Spacetimediagram}(a) and \ref{Spacetimediagram}(b)), the perturbation is weak and not able to spread out over the recirculation zone. Therefore, the unsteady pattern can not sustain or regenerate anymore; it gets carried downstream and the flow becomes globally laminar. However, due to the exerted perturbation at the inlet, a very small perturbation pattern with the bounded \textcolor{black}{($i.e.$ small and quasi constant)} amplitude still exists in the laminar state, which is called weak unsteady pattern (WUP). % WUP is localized in the upstream portion of the flow compare to the reattachement point.
Depending on the values of $\mathcal{A}$ and $Re$, WUP can stay bounded with a small amplitude (for instance, from $t\approx750s$ to $t\approx 1000\ $(s) in figure \ref{Spacetimediagram}(b)), and appears globally as a LS. The flow could experience a transient growth and become a large scale unsteady state (for instance, from $t\approx1000$ in figure \ref{Spacetimediagram}(b)). The resulted state is clearly different from US1 described earlier and will be called as the second unsteady state pattern, hereafter denoted by US2. Comparing to US1, the unsteady pattern of US2 can evolved from an LS, the recirculation zone is mostly preserved (see figure \ref{contourplot}(b)), and the overall amplitude of US2 is smaller than its counterpart in US1. An intermittency behaviour can be observed for the case of $\mathcal{A} \approx 0.0782$ (see figure \ref{Spacetimediagram}(a)), where the unsteady pattern of US2 emerged but decayed after. This pattern will be analysed in more details in the following sections. 
%Moreover, since US2 preserve the recirculation zone, the transition from US2 to LS might be reversible: it's possible to observe these 2 states and the transition within a single simulation. Obviously, if a transformation is reversible, it could present hysteresis behaviour. This property correspond to the first or the second case scenario described in the previous section.

%Differents transition evens is defined: the cases where the flow stays in laminar state (the unsteady pattern at the initialization is not taken into account) until the end of simulation is called CL, the cases where the flow starts with unsteady state and stay at this state until the end is called CT. If a case experience a one way evolution from one state to another (not counting the initial perturbation), it will be labeled as CS1. If a case can evovle in both way, it will be labeled as CS2. 

Other values of Reynolds numbers also show a similar behaviour. The observed states of the flow namely, LS, US1 and US2 are reported in Table \ref{Recap_simu}. Note that while the state US1 can be firmly determined, the boundary between US2 and LS depend on how long we observe the system, since US2 can emerge from LS at very late in the simulation. Therefore, the state reported as LS could be US2 if the observation time is larger, and the boundary can not be firmly determined. Thus, the laminar state in Table \ref{Recap_simu} are reported at least after $t=1000\ $(s). Larger observation time are used as $Re$ approaching the critical value.

%In the table \ref{Recap_simu}, one can see a link between the state (LS,SUS1,SUS2) and the possible transition. This link can be explained based on the relative strength of pertubation: if the pertubation is too strong, it will break the recirculation zone and the turbulence starts immediatly and never experience the laminar state (CT and SUS1). On the other hand, if the perturbation is smaller, the flow try to return to a asymmetric but steady state and in the same time SUS2 waits for emerging. The onset of SUS2 depend on the intial condition and the strength of perturbation. If SUS2 emerge before the flow laminarized, we have state CT and SUS2 (case 9d where $Re=2000$, $\delta=0.1$), but SUS2 could also emerge very late in the simulation, then the flow seem to be completly laminarized and then the instability suddenly emerge. The state SUS2 is self-sustainable in most of our cases (CS1 and SUS2), but there are fews cases where SUS2 is not self-sustainable (case 9b: $Re=2000$ and $\delta=0.0782$), which correspond to CS2 and SUS2. To sumarize, these following couples will be observed in order as the perturbation increase: (LS,CL), (CS2,SUS2), (CS1,SUS2), (CT,SUS2) and (CT,SUS1). 

In the figure \ref{Transition_boundary}, the boundaries of LS and US2 are shown. The first observation is the border of transition follows a power law of $\mathcal{A} \sim Re^{-3}$, which is much steeper than the value of $-0.006$ that is reported by Sanmiguel-Rojas \& Mullin\cite{rojas2012}. \textcolor{black}{Meanwhile, the figure provided by the authors shows a completly different powerlaw, which estimated around $-11.2$. This important results yet confusing deserve more attendtion. Moreover, the perturbation method used by \cite{rojas2012} is a tilt disturbance, $i.e.$ shifting the inlet velocity profile upward (in X direction). This creates a discontinuity in the inlet and at the wall, which causes an intense shear in upstream flow. The effects of the shear generated could depend on the mesh resolution as well as on how the solver interpolate the discontinuity; but not only the physical parameters. A recent experimental study from \cite{Lebon2018} show a powerlaw of $ Re^{-2.3}$, which is much closer to our results. }

The second observation is that the band of parameters that exhibits US2 is very narrow, and it situates in the small border between turbulence state US1 and the laminar state LS. Therefore, it plays an important role in understanding of the transition mechanism, especially when the $Re$ and/or $\mathcal{A}$ parameters evolves gradually. The state US2 is new compared to the state US1 which is well documented, so in the following section, this state will be analysed in more details. 

\section{Transient growth of weak unsteady pattern}

In this section, the WUP state and the transition from LS to US2  will be further investigated. The smoothness of this transition make it particularly interesting to study, since it usually happens too fast to be captured.

\begin{figure}
\centering
\includegraphics[scale=1]{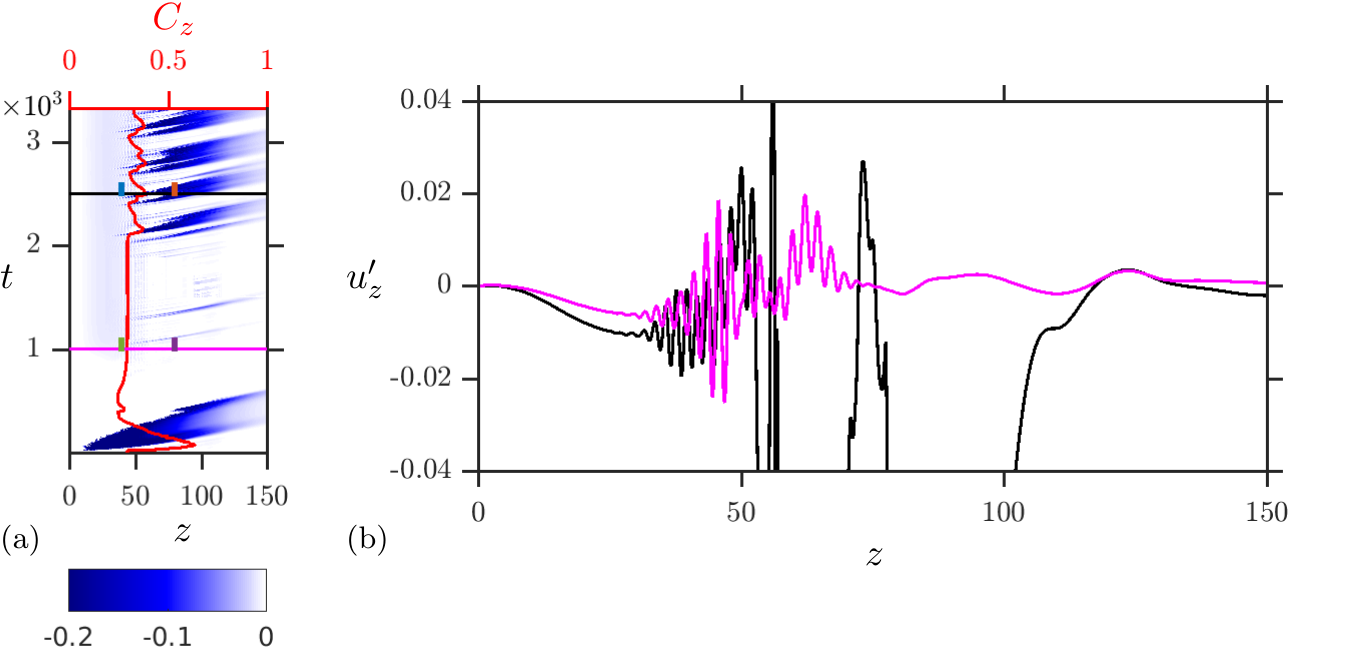}
\caption{Analysis on case 5a: subplot (a) - drag coefficient ({\color{red}\rule{0.5cm}{2pt}}) superimposed on the space-time diagram. Subplot (b) variation  of $u'_z(0,0,z,t)$ in z direction for $t=1000\ $ (LS {\color{magenta}\rule{0.5cm}{2pt}}) and $t=2500\ $ (US2 {\color{black}\rule{0.5cm}{2pt}}). \label{Case_5a}}
\end{figure}

Figure \ref{Case_5a}(a) presents the flow's space-time diagram for $Re=1360$ and $\mathcal{A}=0.2$, corresponding to case 5a in Table \ref{Recap_simu}.
%The case (5a) is realized with $Re=1360$ the initial state of case (5a), with $Re=1350$. 
This case is selected because, it contains unsteady regions and it is the closest case to an LS scenario. The flow stays for a long period in LS, and the transient growth arises around $t\approx 2050$.% We expect the transition to happen more gradually. 

The time evolution of the system can be divided into three phases. The initial phase when the flow adapts to the new parameters.  Before $t<600$, an un-physical instability develops and decays downstream. After the initial phase, the flow experiences a long lasting laminar state, for $600\ <t<2000\ $, with a bounded unsteady pattern WUP. Then, at $t>2000\ $ the flow \textcolor{black}{ shows, for this particular case, an intermittent pattern between laminar state (LS) and turbulent state (US2). Note that For slightly higher $Re$, the turbulent state US2 is self sustained,  the intermittent pattern ceases.}

\begin{figure}
\centering
\includegraphics[scale=1,trim={0.0cm 0cm 0cm 0cm},clip]{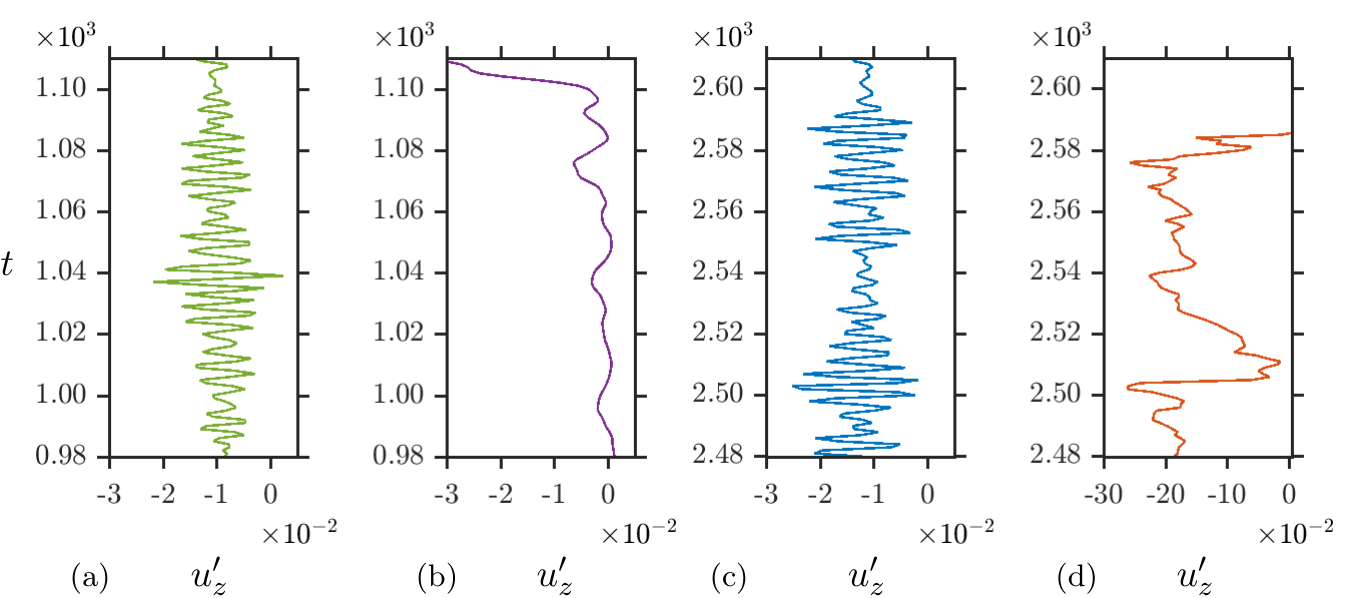}
\caption{Time signal of $u'_z(0,0,z,t)$ in laminar state recorded for $z=40d$ (a {\color{matlab5}\rule{0.5cm}{2pt}}) and $z=80d$ (b {\color{matlab4}\rule{0.5cm}{2pt}}). Time signal of  $u'_z(0,0,z,t)$ in US2 recorded for $z=40d$ (d {\color{matlab1}\rule{0.5cm}{2pt}}) and $z=80d$ (e {\color{matlab2}\rule{0.5cm}{2pt}}). The position of plots in this figure are marked with the corresponding colour in figure (\ref{Case_5a}a).\label{Case_5abis}}
\end{figure}

To take a closer look at WUP, two time snapshots (figure~\ref{Case_5a}b) of $u'_z$ along the streamwise direction are taken at $t=1000\ $ (LS) and at $t=2500\ $(US2). The flow contains  WUP even in the laminar stage. Additionally, three distinct space regions, which two of them are identical in both LS and US2, can be seen.

The first zone is a steady region for $0<z<20d$, where the flow is completely \textcolor{black}{laminar and time independent}. 
The second zone extent from the first zone to the reattachment point, approximately at $20<z<50d$, where WUP show a clear wavy pattern in space with the wave length of $\lambda_{WUP}\approx 2.34d$. 

%\textcolor{blue}{Has it a meaning? (see for instance Betchov \& Szewczyk 1963)  Orr-Zommerfeld equation gives $\lambda=2d$ in case of shear layer instability} 

Figure \ref{Case_5abis} presents two time signals of streamwise velocity fluctuations recorded respectively at $z=40d$ and $z=80d$ over the pipe centerline ($i.e.$ $x=y=0$). The selected time signals in this zone (see figures \ref{Case_5abis}(a) and \ref{Case_5abis}(c)) have a similar noise's amplitude $u'_{WUP}/\overline{U}\approx0.1$ and frequency $f_{WUP}\approx0.2 Hz$. This witnesses the similarities in the second zone for WUP in both LS and US2.%%%%%%%%%%

In the third and last zone, $i.e.$ the region after the reattachment point, WUP transit into larger scale dynamics. In this region, the low frequency disturbances start to merge (see also figure \ref{Case_5abis}(b) purple line), with larger wave length in the case of LS. If the parameters $Re$ and $\delta$ are sufficiently high, the unsteady pattern in third zone can be amplified into a strongly time dependent \textcolor{black}{and non linear pattern US2}, see also figure \ref{Case_5abis}(d). 

As mentioned earlier, the second and the third zones are situated before and after the reattachment point. \textcolor{black}{This point, beside the mathematical definition, can also be visually determined from the contour plot of the streamwise velocity at level $v_z=0$ }(see figure \ref{contourplot}). The boundary between the second and the third zones is not sharp because the WUP shows organized structures which slowly decays as they crossed the border. This position indicates an important change in the flow dynamics.

The before mentioned observations are in agreement with the transient growth analysis performed by  Cantwell et al.~\cite{Cantwell2010} in the sense that the disturbance generated at the inlet grows while traveling downstream, mainly because of its interactions with the recirculation bubble. However, the authors would like to highlight some differences in our methodology and results when compared to \cite{Cantwell2010}. Contrary to linear stability analysis with infinitesimal perturbations of \cite{Cantwell2010}, DNS with finite amplitude perturbations are performed in the present work. In linear stability analysis, perturbations can either decay or grow, but in our case, perturbations could stay bounded for a very long time and then suddenly grow. The nonlinear interactions can also be observed in the DNS data, especially when there are interactions between the unsteady patterns and the recirculation bubble.

%The unsteady pattern emerge at around $t\approx2000s$. 
%%%%%%%%%%%%%%%%%%%

%In the following study, more detail on the mechanism of the transition will be presented. 

The WUP pattern can be also extracted from the flow by using a Reynolds averaging technique on the LS phase. The base flow is approximated by taking the average fields during the LS (from $t=1500\ $(s)  to $t=1800\ $(s)): 
\begin{equation}
\overline{\mathbf{u}}(x,y,z)=\left<{\mathbf{u}}(x,y,z,t)\right>_{1500<t<1800}.
\end{equation}
Then the unsteady pattern can be extracted using:
\begin{equation}
\mathbf{u'}(x,y,z,t)=\mathbf{u}(x,y,z,t)-\overline{\mathbf{u}}(x,y,z).
\label{uf}
\end{equation}

Since the amplitude of $\mathbf{u'}(x,y,z,t)$ is small and stay bounded in laminar phase, it could be treated as a perturbation and qualitatively compared with linear stability analysis. To further compare our results with \cite{Cantwell2010}, the coordinate system is converted  from Cartesian ($\mathbf{u'}(x,y,z,t)$) to the cylindrical ($\mathbf{u'}(r,\theta,z,t)$). Then, the Fourier transform of $\mathbf{u'}(r,\theta,z,t)$ in azimuthal direction, noted as $\widehat{\mathbf{u'}}$, is computed:
\begin{equation}
\widehat{\mathbf{u'}}(r,k_{\theta},z,t)=\frac{1}{2\pi}\int\limits_0^{2\pi}\mathbf{u'}(r,\theta,z,t)e^{-2i\pi\theta k_{\theta}}d\theta,
\end{equation}
where, the azimuthal modes energy are computed point-wise as:
\begin{equation}
\left( \begin{array}{ccc}
e_z(r,{k_{\theta}},z,t) \\
e_y(r,{k_{\theta}},z,t) \\
e_z(r,{k_{\theta}},z,t)\end{array} \right)=\frac{1}{2}\left( \begin{array}{ccc}
|\widehat{u'_x}(r,k_{\theta},z,t)|^2\\
|\widehat{u'_y}(r,k_{\theta},z,t)|^2 \\
|\widehat{u'_z}(r,k_{\theta},z,t)|^2\end{array} \right),
\end{equation}
and the total energy of each azimuthal modes is the sum of all three directions: 
\begin{equation}
e_{total}(r,{k_{\theta}},z,t)=e_x(r,{k_{\theta}},z,t)+e_y(r,{k_{\theta}},z,t)+e_z(r,{k_{\theta}},z,t).
\end{equation}

%We could observe that  $e_{total}$ is mainly contributed by $e_z$.  To prouve the domiance of fluctuation in streamwise direction, we compute the lower bounded value of contribution  in percentage of $e_z$ toward $e_{total}$ over all $r$ and $z$ value:
%\begin{equation}
%P(k_{\theta},t)=\min\limits_{z,r}\left(\frac{e_z(r,k_{\theta},z,t)}{e_{total}(r,k_{\theta},z,t)}\right)
%\end{equation}
%
%In figure .. we plot the time evolution of $P(k_{\theta},t)$ for the most relevant modes ($ k_{\theta}  \in \{0,1,2,3\}$). Therefore, we can only focus on the fluctuation in streamwise direction $u'_z$. 

From the linear stability point of view, the perturbation $\mathbf{u'}$ \textcolor{black}{could be decomposed into grow and decay modes}. However, in our case, its magnitude stays bounded and fluctuates for a very long time, before \textcolor{black}{the transient growth started and US2 emerge. It suggests there should be a very slow growing mode or new perturbation modes are introduced. This is what we investigated in this section.} The observation of the time evolution of all azimuthal modes (over all $z$, $r$ and $k_{\theta}$) reveals  no steady increasing of energy. Instead, in addition to the regular WUP,  a perturbation appears in the steady zone ($z<20d$) and smoothly evolves into turbulence. 

\begin{figure}
\centering
\includegraphics[scale=1,trim={0.cm 0cm 0cm 0cm},clip]{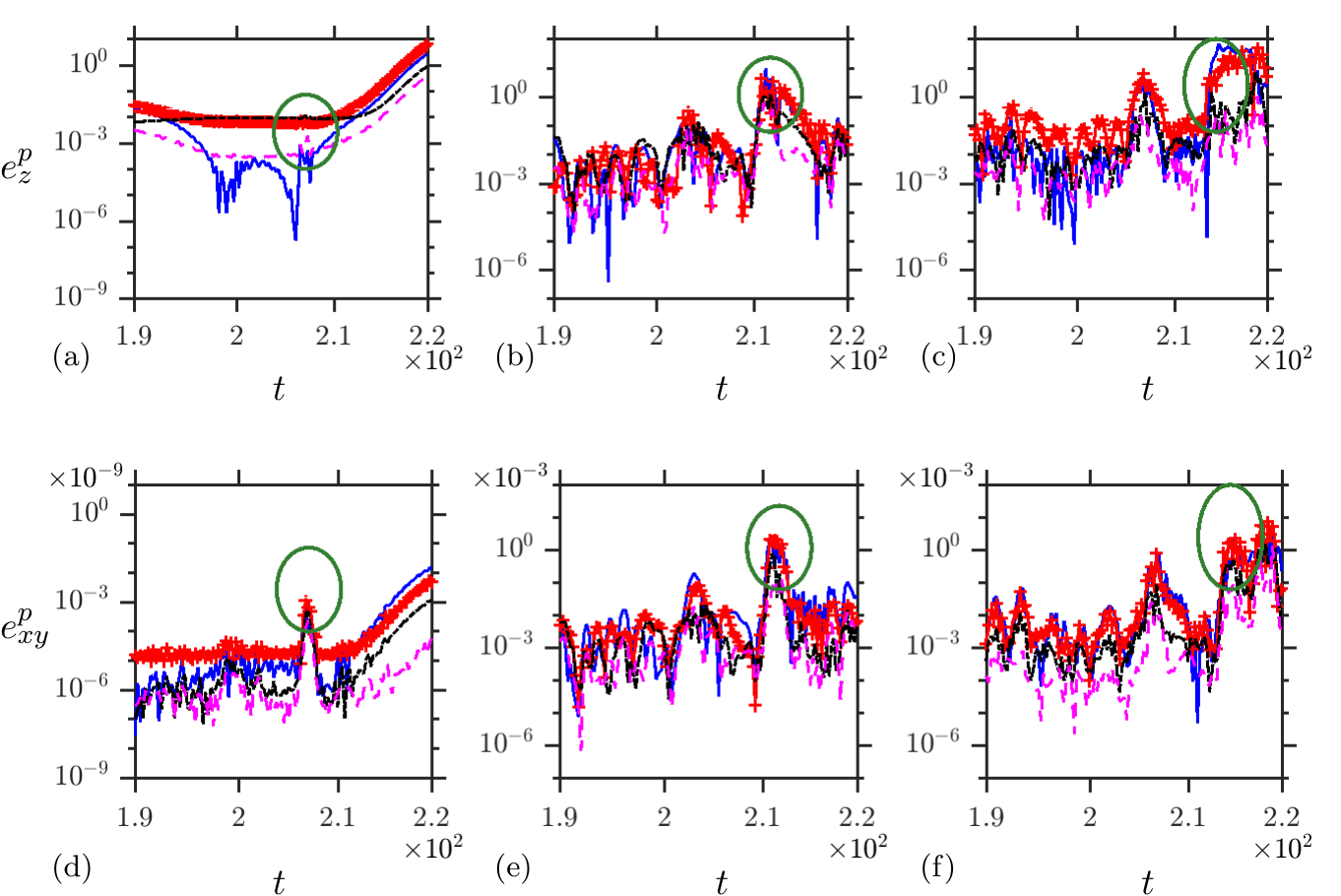}
\caption{Time evolution of amplitude of the most relevant azimuthal modes at $r=0.25d$. The colour code is: blue line for the mode 0, red (thick) line for the mode 1, black line for mode 2, and magenta line for mode 3. The 3 plots in upper row (a,b,c) show the energy of fluctuation pattern in streamwise direction ($e_z$), and the 3 plots in the lower row (d,e,f) show the energy of fluctuation pattern in spanwise direction ($e_{xy}=e_x+e_y$). The 3 columns correspond to respectively $z=17d$ (figure (a),(d)), $z=47d$ (figure b,e) and $z=67d$ (figure (c),(f)). The elipse annotation high light the spotted peak of energy.
\label{Time_evolution_modes}}
\end{figure}

A closer look into the time evolution of the perturbation amplitude in the vicinity of transition ($1900<t<2200$) is shown in figure \ref{Time_evolution_modes} for an arbitrary value of $r$ (here $r=0.25d$) and three different streamwise positions. \textcolor{black}{For clarity, only one value of $r$ is shown, the other values of $r$ exhibit similar behaviors}. In this figure, the fluctuations of energy in streamwise direction ($e_z(r,{k_{\theta}},z,t)$) or spanwise direction ($e_x(r,{k_{\theta}},z,t)+e_y(r,{k_{\theta}},z,t)$) are recorded as a signal in time for given values of $z$, $r$ and $k_{\theta}$. From these signals, a local peak in the time evolution can be noticed that is located close to the transition moment. This peak can be recorded as a time instance $t_p$ corresponding to the maximum of the energy signal. The value of collected $t_p$ will be a function of spatial positions $r$, $z$ and the mode $k_{\theta}$. 

By comparing the fluctuation patterns, it is found that fluctuations in streamwise direction are around two orders of magnitude stronger than the sum of energy fluctuations in the two other spanwise directions \textcolor{black}{in the laminar zone} (first zone: $z<20d$). However, the gap in the energy levels gets closer at the transition point in the second zone: $z\approx50d$ and the non linear region, in the third zone: $z>60d$. 
Another observation is that the position of the peaks mainly depends on the streamwise position $z$ and seems to be independent of spanwise variables ($r$ and $k_{\theta}$). For $z\approx 17d$, peaks start to have significant value  around $t=2080\ $(s). 

For a given position, the pattern of the peaks, $u^{p}_z$, can be extracted by subtracting the velocity fields at instance before the peak ($t=2070\ $(s)) from the instance at the peaks ($t_p=2080\ $ (s))
\begin{equation}
\label{definePP}
u^{p}_{z}(x,y,z=17d)=u_z(x,y,z=17d,t=t_p=2080\ (s))-u_z(x,y,z=17d,t=2070\ (s)).
\end{equation}

\begin{figure}
\centering
\includegraphics[scale=1,trim={1.5cm 0.2cm 0.2cm 0.2cm},clip]{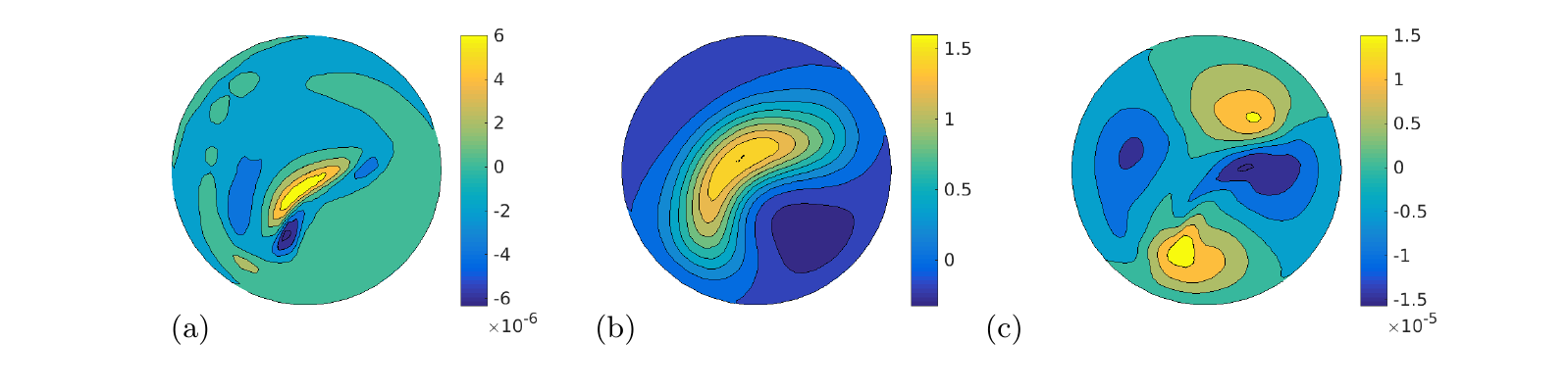}
\caption{Surface plot in spanwise plan, slide at $z=17d$ of: figure (a) pattern that growth rapidly $u^p_z(x,y,z=17d)$ (see definition in equation (\ref{definePP}) , figure (b) mean flow: $\overline{u}_z(x,y,z=17d)$, figure (c) regular unsteady pattern that stay bounded over time: ${u'_z}(x,y,z=17,t=2070\ (s))$\label{Slide_z}}
\vspace{-0.2cm}
\end{figure}

Figure \ref{Slide_z} presents the contour plots of different streamwise velocity patterns. It is found that the most intense perturbation in the peak pattern $u^{p}_{z}$ (figure \ref{Slide_z}(a)) appears close to the strongest shear rate position in the streamwise velocity mean profile $\overline{u}_{z}$ (figure \ref{Slide_z}(b)). This suggests  the peak pattern is a shear instability and is independent of the regular unsteady pattern $u'_{z}$. The regular unsteady pattern (figure \ref{Slide_z}(c); see also equation \ref{uf}) is the fluctuation collected before the peak emerge. The latter structure is completely different from both the peak pattern and the mean flow profile structures and the fact that it stays bounded for a large period of time suggests that it would not evolve into transition mechanism.

%As discused earlier, $RP$ is an unsteady pattern that is generated by the interaction of radial flux with the mean velocity profile, it will stay bounded

\begin{figure}
\centering
\includegraphics[scale=1,trim={0cm 0cm 0cm 0cm},clip]{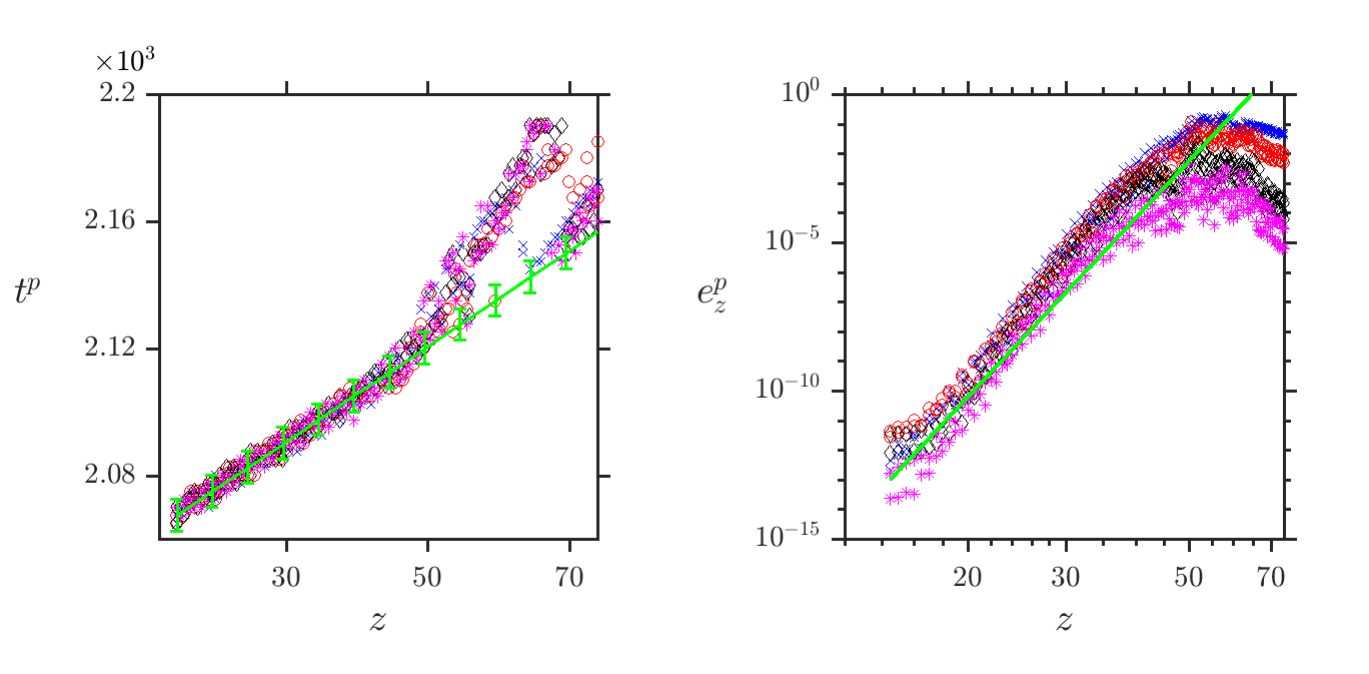}
\caption{Tracking position of peak: figure (a) peak position in $z$ versus its position in time, with the scaling in green line: $t=z/V_0+constant$, figure (b) peak position versus its amplitude for first 4 modes: mode 0 ($\textcolor{blue}{\times}$), mode 1($\textcolor{darkorange}{o}$), mode 2($\textcolor{black}{\diamond}$), mode 3($\textcolor{magenta}{*}$), with the scaling in green line $e^{p}_z=constant . z^{20}$. }
\label{Tracking_peak}
\vspace{-0.4cm}
\end{figure}

By fixing $r$ at any arbitrary value (here we chose $r=0.25d$) and focusing on the first 4 modes ($k_{\theta}\in\{0,1,2,3\}$), one could notice that the position of local peak evolves smoothly and linearly in the streamwise direction in the range of $10d<z<50d$. This suggests that the peak is an perturbation pattern that appears in addition to the regular pattern, it gets carried downstream by the main flow and meanwhile amplified. In figure \ref{Tracking_peak}(a), the tracking of the local peaks in space ($z$) and time ($t$) are shown (all the value of $r$ are superposed), where a linear fit of $z/V_0+constant$ with $V_0\approx 0.667$ is found for \textcolor{black}{the same region.}

The error of the tracking process mainly comes from the the output frequency of the data and the fluctuation in $z$ direction. All the peaks found in the range $10d<z<40d$ are contained within the linear fit with the error bar of $\pm 2\delta_t$ where $\delta_t$ is the distance traveled by the peak between two consecutive outputs. It is interesting to mention that the peak evolution tracking in space ($z$) and in time ($t$) collapse for all the values of spanwise variables ($r$ and $k_{\theta}$). This indicates  the perturbation pattern is carried downstream by the same velocity $V_0$, for all $\forall r$ and $\forall k_{\theta}$. $V_0$ is expected to be close to the velocity profile where the perturbation is the most intense. The disturbance is generated punctually at $(x_0,y_0,z_0)$, gets carried downstream with the mean velocity in the same position $\overline{u}_z(x_0,y_0,z_0)$ and then distributed in 2 spanwise directions. The maximum value of $u^{p}_{z}(x,y)$ corresponds to $x=-0.0097d$ and $y=-0.1221d$. At this position, the mean flow has the value of $0.58$, which is roughly close to $V_0$.

When the pattern of $u^{p}_{z}$ is carried out of zone 2 and enters zone 3 (at $z\approx50d$), its velocity is slowed down. From this point, the non linearity starts to emerge, the peak is diffused and merge with another peaks that emerged earlier. As a consequence the localized peak in time is transformed into a plateau. In figure \ref{Tracking_peak}(b), the magnitude of the peaks over time are plotted for the first 4 azimuthal modes. The growth of all the modes follow a well defined power law, until $z\approx 50 d$ where they start to saturate.

Finally, we believe the chaotic behaviours that emerged in the third zone ($z>50d$) are the same convective instabilities that are generated by the shear in the first zone ($z\approx17d)$. The generation and evolution of the shear instability seems to be independent of the existence of other unsteady behaviours. It is noted that the phenomenon of the convective instability is also observed in the numerical simulation of flow with backward-facing step \cite{blackburn_barkley_sherwin_2008}.

%It seem that the peak is a punctual disturbance (PD) on top of the regular disturbance (RD), its start to have significant value at $z=15d$. 

\section{Hysteresis}
\begin{figure}
\centering
\includegraphics[scale=1,trim={0cm 1cm 0cm 0cm},clip]{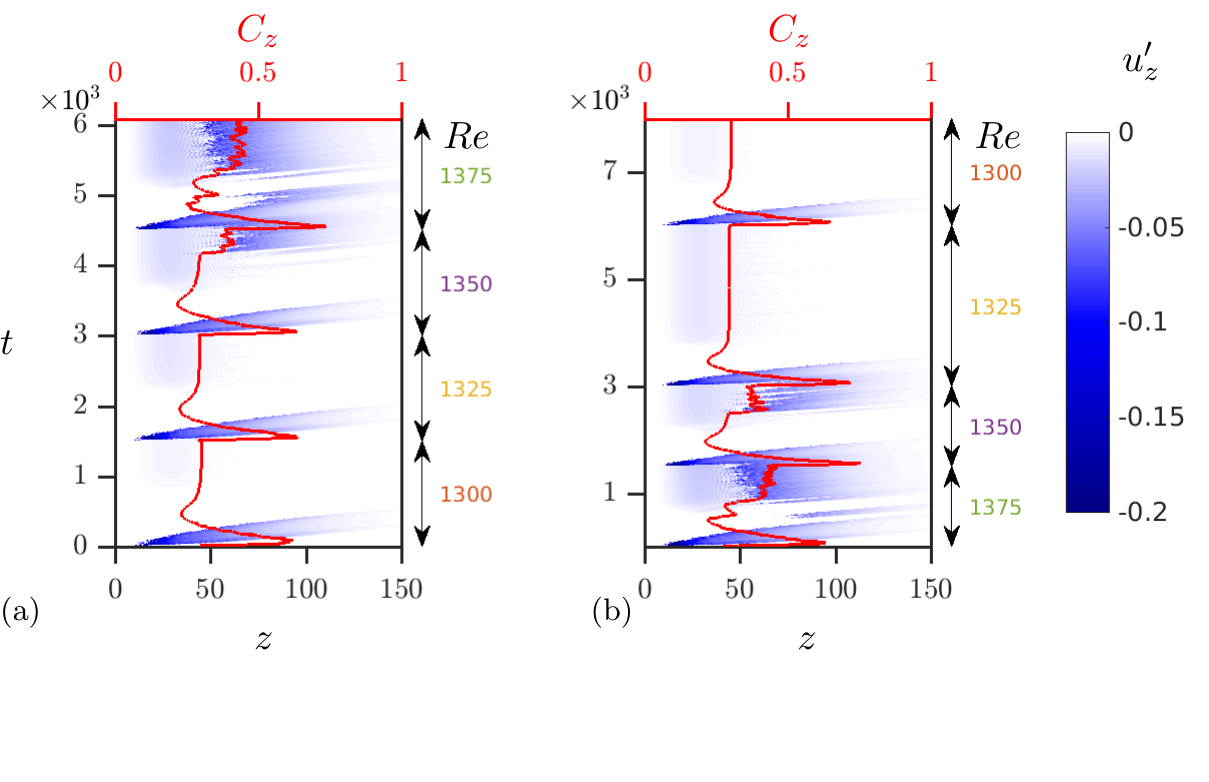}
\caption{Hysteresis study of transitional flows. ({\it a}) Space-time diagrams of the increasing $Re$, (cases 3a, 4a, 6a and 7a) and ({\it b}) the decreasing $Re$, (cases 7b, 6b, 4b and 3b) branches. The diagrams also indicate the value of $C_z$ as a (red) continuous line on the top and the corresponding value of $Re$, changes are indicated on the right. }
\label{Hysteresis}
\end{figure}

\begin{figure}
\centering
\includegraphics[scale=1,trim={0cm 1cm 0cm 0cm},clip]{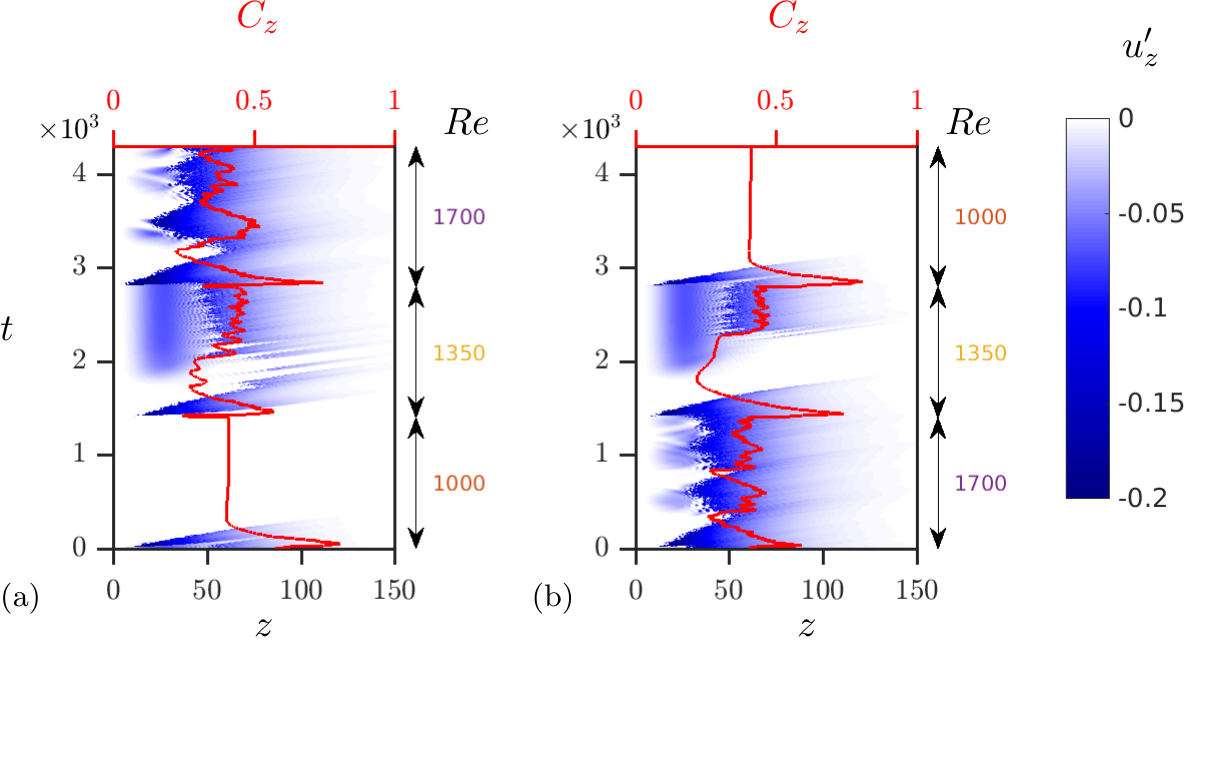}
\caption{Hysteresis study of transitional flows. ({\it a}) Space-time diagrams of the increasing $Re$ (cases L1d,L2d and L3d) and ({\it b}) the decreasing $Re$ (cases L3d,L2i,L1i) branches. The diagrams also indicate the value of $C_z$ as a (red) continuous line on the top and the corresponding value of $Re$ changes are indicated on the right.}
\label{Hysteresis3}
\end{figure}

\begin{figure}
\centering
\includegraphics[scale=1,trim={0cm 0cm 0cm 0cm},clip]{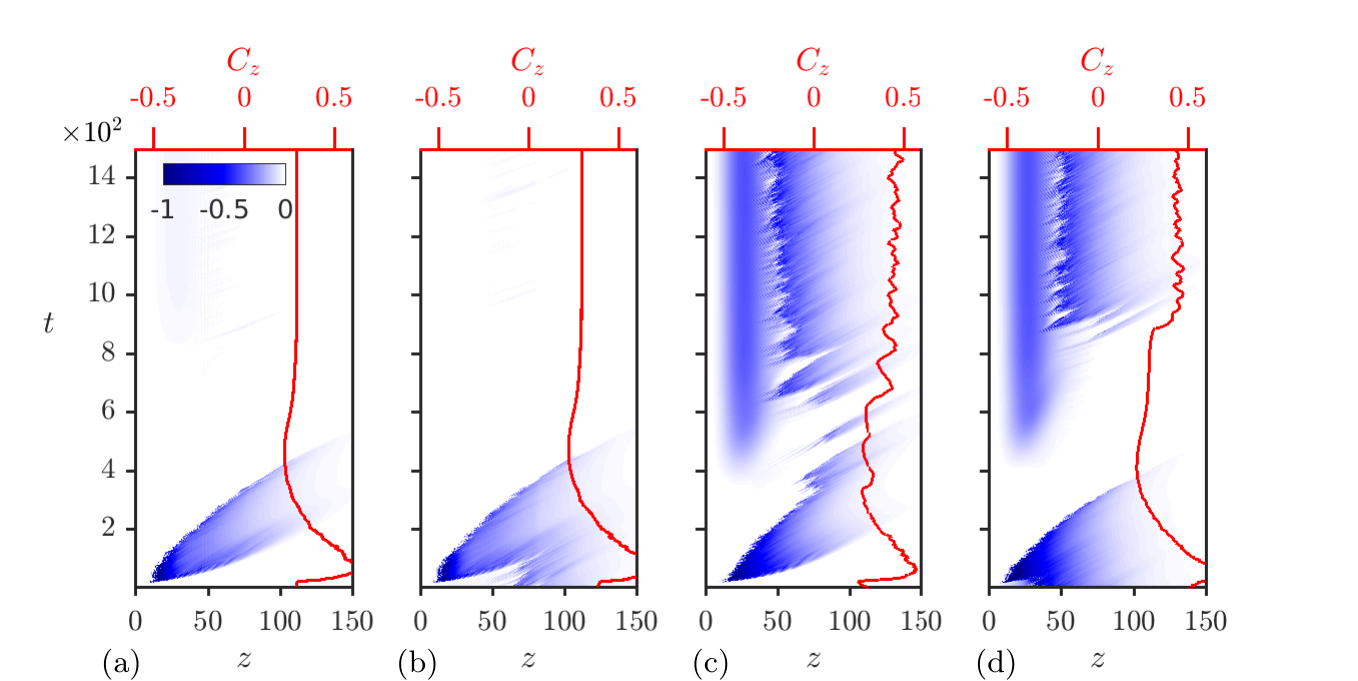}
\caption{Space-time diagrams for $Re= 1350$ and $\mathcal{A}=0.2$ with different initial conditions, respectively from left to right, ({\it a}) $Re = 1325$; ({\it b}) $Re = 1375 $; ({\it c}) $Re= 1000$, and ({\it d}) $Re= 1700$}
\label{Hysteresis2}
\vspace{-0.4cm}
\end{figure}

\begin{figure}
\centering
\includegraphics[scale=1,trim={0cm 1cm 0cm 0cm},clip]{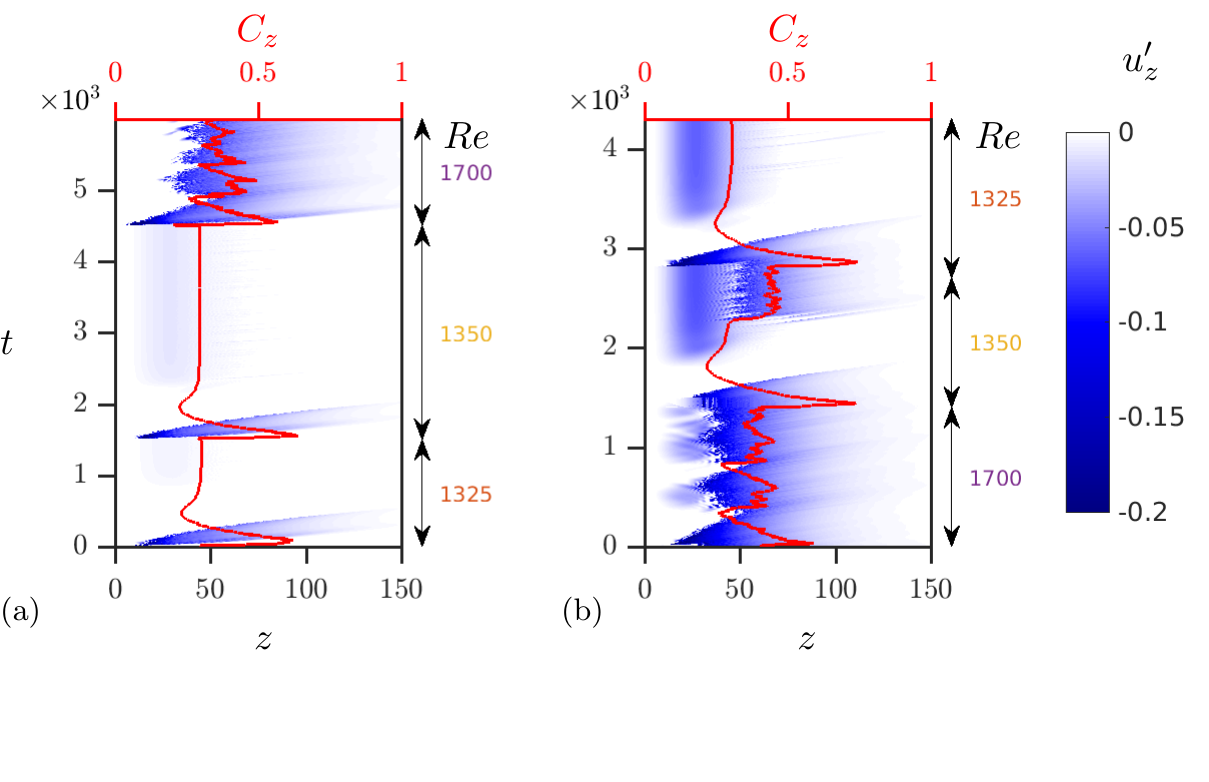}
\caption{Hysteresis study of transitional flows. ({\it a}) Space-time diagrams of the increasing $Re$ (cases L1d,L2d,L3dbis) and ({\it b}) the decreasing $Re$ (cases 3a, 4a and 10a) branches. The diagrams also indicate the value of $C_z$ as a (red) continuous line on the top and the corresponding value of $Re$ changes are indicated on the right.  ($\mathcal{A}=0.5$)}
\label{Hysteresis4}
\end{figure}

The hysteresis can appears when two flow solutions can exist for the same $Re$. Therefore, the initial conditions and mainly the parameters of the disturbance control the appearance of the two solutions. To test the hysteresis behaviour, two branches of the simulations with $Re$ increasing and $Re$ decreasing are investigated. The $Re$ increasing branch starts from laminar flow, whereas the $Re$ decreasing branch begins from an unsteady state, here is US2. This approach was also considered by  \citet{rojas2012} using a tilt disturbance of amplitude $\delta$. Depending on $\delta$ and $Re$ a domain of hysteresis was observed by the same authors. Specifically, for $\delta=0.001$, the coexistence region was reported for $1475<Re<1850$. Moreover, they found that the hysteresis region grows as $\delta$ decreases. It is not possible to directly compare the effect of the tilt disturbance with our vortex disturbance because these perturbations are of different nature: the vortex disturbance perturbation introduces rotation, whereas the tilt disturbance introduces a translation to the flow that does not obey non-slip velocity condition at the wall and can therefore be considered unphysical. However, the following results discuss the universality of the hysteresis behaviour.

A series of eight simulations are performed, with a fixed amplitude of vortex perturbation, $\mathcal{A}=0.2$. The initial condition is case 3a with $Re= 1300$. When the simulation time reached $1500s$, the final state is analysed and used as the initial condition for the next run at a higher $Re$. Then, $Re$, is increased with step  of 25, up to $Re=1400$ (case 4a, 5a, 7a, 8a), as depicted in the space-time diagrams of figure \ref{Hysteresis}({\it a}). In figure \ref{Hysteresis}({\it b}), the decreasing $Re$ branch is initialised with a laminar state (case 3a), then  $Re$ is directly increased to 1400 (case 8b) and the decreasing path down to 1300 with a step of 25 (case 7b, 5b, 4b). The simulations from the increasing and decreasing paths are compared for the same $Re$ the results are presented in figure \ref{Hysteresis}({\it a}) and \ref{Hysteresis}({\it b}) and show minor changes.
Looking at the drag coefficient also represented in figure \ref{Hysteresis}({\it a}) and ({\it b}), every change in $Re$ initiates a peak or a transient increase of $C_z$ corresponding to the turbulent spot that propagates and decays downstream. As seen no hysteresis is found while following this standard procedure.

Additional simulations using different $Re$ steps and a larger range of $Re$, were performed. The decreasing branch started with $Re=2000$. Then $Re$, is decreased to 1700, 1350 and 1000 (simulation L1d, L2d, L3d). The results is compared with the increasing branch which start from $Re=1000$ then increase to $1350$, then $1700$ (simulation L3d, L2i, L1i). Again, the results are represented in the form of space-time diagrams, in figure \ref{Hysteresis3}({\it a}) and \ref{Hysteresis3}({\it b}). The data show minor differences in the space-time behaviour. However, this time the drag $C_z$, suggests a small loop of hysteresis. 

When comparing two series of simulation, one could notice: in the first loops, with small steps of $Re$ ($\Delta Re=25$), the two cases with $Re=1350$ are laminar in both increasing and decreasing branches. Whereas, in the second series, with larger steps of $Re$ ($\Delta Re=350$), the two cases with same $Re=1350$ show US2. The space-time diagram of 4 cases with $Re=1350$ are extracted in the two series and show in figure \ref{Hysteresis2}. The different behaviour of the same $Re$ suggest that larger steps of $Re$ could eventually trigger the unsteady behaviours and potentially hysteresis loop sooner than smaller steps.

To quantify the extent of hysteresis behaviour, we first define the integral of $C_z$ over $Re$ for both increasing and decreasing branches, 
\begin{equation}
 S=\int C_z(Re)dRe ,
 \label{define_Hysteresis1} 
 \end{equation}
 and then the relative difference is used to quantify the hysteresis behaviour:
\begin{equation}
 H=\frac{\Delta S}{\overline{S}}.
 \label{define_Hysteresis2} 
 \end{equation}
 
Here $\Delta S$ is the difference between 2 branches and $\overline{S}$ is the mean value. It is noted that in order to observe the hysteresis phenomenon clearly, a specific procedure needs to be implemented. In the decreasing branch, the variation of $Re$ should be large enough to avoid the transformation from US1 to US2. On the other hand, in the increasing branch, the variation of $Re$ should be small enough to keep  the flow laminar. Such loops is presented in figure \ref{Hysteresis4}, with the decreasing branch initiates at $Re = 2000$, which is decreased consecutively to 1700, 1350 and 1325 (simulation L1d, L2d, L3dbis). The increasing branch is initiated at $Re =  1300 $ and is then increased consecutively to 1325, 1350 and 1700 (simulation 3a, 4a, 10a). Based on the criteria defined in equation \ref{define_Hysteresis2}, this procedure lead to a hysteresis of $H=27.87\%$ compare to the two previous loops with the measure of hysteresis are $H=1.15\%$ and $H=0.75\%$ respectively.

\begin{figure}
\centering
\includegraphics[scale=1,trim={0.2cm 0.3cm 0cm 0cm},clip]{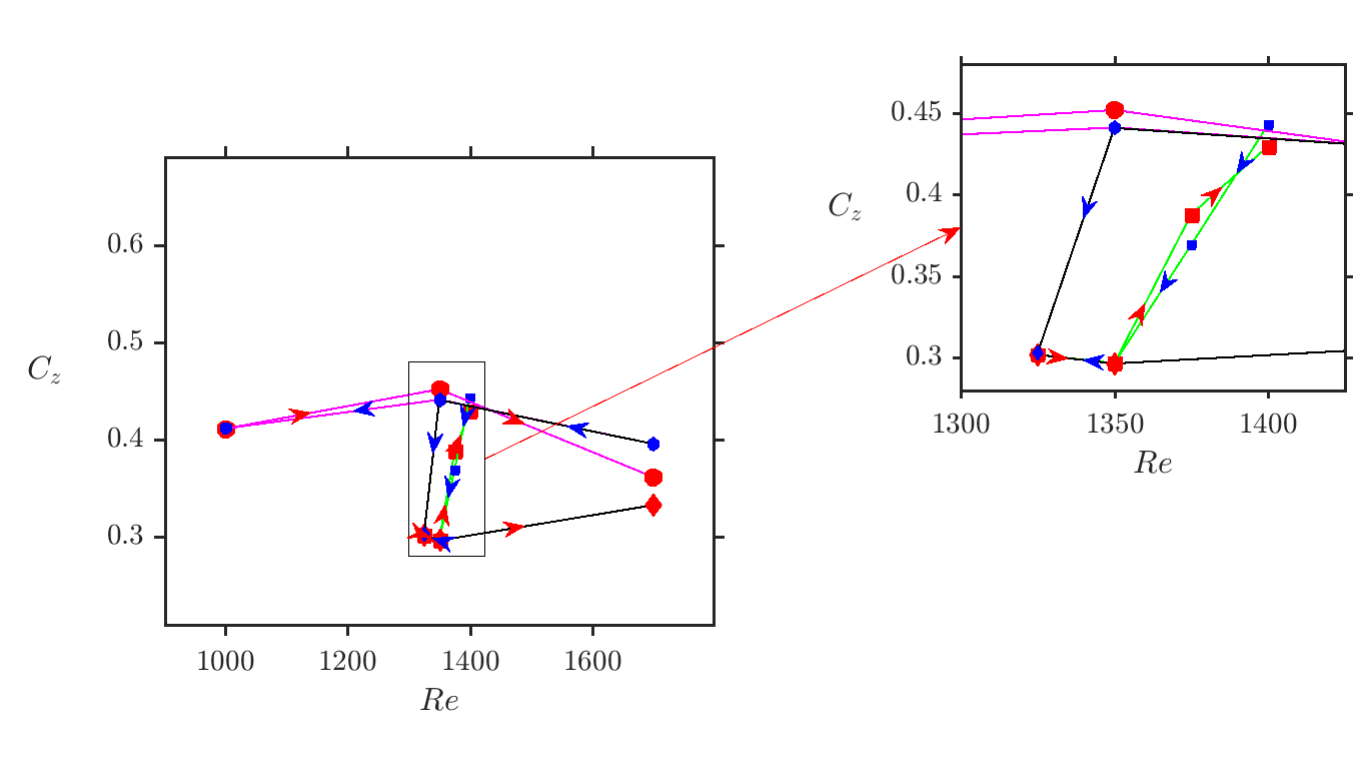}
\caption{Hysteresis loop of turbulent transition : $C_z$ as a function of $Re$. The left hand side plot is the global view and the right hand-side plot show a zoom into rectangle region of the right hand-side plot. The blue line is a fit of $C_z$ using cases with several $Re$ and $\mathcal{A}=0$, its expression: $\frac{900}{Re}-0.57$. 3 sequences of simulations are presented: the cases shown in figure \ref{Hysteresis} in green line with square markers, the cases shown in figure  \ref{Hysteresis3} in magenta lines with circular markers and the cases shown in figure  \ref{Hysteresis4} in black line with diamond markers. In each sequence, the simulations in increasing branch are differentiate with the ones in decreasing branch by the arrows showing the direction of variation of $Re$ . Additionally, the increasing branch are presented with slightly bigger red marker and red arrows, whereas the one in decreasing branch are highlighted with smaller blue marker and blue arrows. }
\label{GatherCz}
\end{figure}
In figure \ref{GatherCz}, the $C_z$ from the last $100$ seconds of each case is plotted against $Re$. The laminar states, $Re= 1325$ and $Re = 1350$, lead to $C_z$ with almost the same value within 0.04\%.  The unsteady state at $Re = 1375$ and $Re = 1400$ have slightly different final value of $C_z$ within 3\% because of the unsteady nature of the flow.
It is noted that the systematic study of all possible steps with extremely long time scales would be a tedious investigation thus beyond the scope of this study.

\section{Conclusion}
In the subcritical $Re$, range studied here, $1350\lesssim Re\lesssim 2000$, it was shown that the flow through a sudden expansion in a circular pipe perturbed by a finite amplitude and a constant perturbation could be forced into unsteady states. With a strong enough inlet disturbance, that can break the recirculation region, the unsteady pattern gains the energy during the process and becomes sustainable even when the perturbation is removed. If the perturbation is not strong enough, the flow might return to its laminar state, and then develop an unsteady pattern with a scenario similar to  a linear growth process described by \citet{Cantwell2010}, in which an infinitesimal upstream disturbance can be amplified while being carried downstream, when it reaches certain threshold, it can trigger the instability. This second type of instability exists only within a very narrow band of flow parameters, and it can start emerging at later time. However, it is a state that the flow must cross if it evolves gradually from a laminar state to a turbulence state.

This marginal unsteady state has several distinguish feature compared to the fully developed turbulent state: it does not exhibit hysteresis behaviour at large observation time. With an appropriate parameters (small $Re$ and $\mathcal{A}$), it could decay and the flow can revisit the laminar state or it could self-sustain and the unsteady patch stay at the end of the recirculation zone, it generally has weaker amplitude than the first unsteady state. 
Either way, the time and space signal for either case, $i.e.$ LS or US2, have some similarities. First they show a steady time-independent region right after the expansion at $0<z<20d$; and second  they have a very well organised weakly unsteady patterns, so-called WUP, till the region before the reattachment zone ($i.e.$ $20d<z<50d$). However, after the reattachment zone, $i.e$ $z>50d$, the low frequency disturbances can emerge and create a still bounded disturbance with larger wave length in the case of LS or can be amplified into a strongly time dependent pattern in the case of US2. The perturbation amplitude required for the first instability to emerge, seems to scale with the Reynolds number as a power law of -3, which is much steeper than the previous numerical estimation of \cite{rojas2012}, who showed a power of -0.006.

Furthermore, it was found that the velocity fluctuations in the streamwise direction are dominant ($i.e.$ about two order of magnitude larger) in the first two before-mentioned regions when compared to two other spanwise directions. However, this gap in the velocity fluctuation energy content gets closer after the transition point where the nonlinearities arise. Additionally, a peak fluctuation was observed which was independent of the spanwise variables. By comparing its most intense position to the shear rate map, it was shown that this peak pattern is a shear instability and appears on top of the regular unsteady pattern $u'_z$, where it carried downstream with constant speed and its amplitude grows exponentially ($e^{p}_z \sim z^{n}$).

Finally, a hysteresis quantification procedure was established to clarify the ambiguities in the recent literature. It is found that for fairly small change, increasing or decreasing, in the Reynolds number and sufficiently large computational time, there exists no hysteresis unlike what is reported in \cite{rojas2012}. However, if the change between the initial and the final quasi-steady flow is large enough, for instance by suddenly dropping the Reynolds number by a significant value, the turbulent flow inside the sudden expansion pipe can be self-sustained. For such cases, hysteresis behaviours are found within the reported physical time in the manuscript.

\begin{acknowledgments}

This work was granted access to HPC resources of IDRIS under the allocation 2017-100752 made by GENCI (Grand Equipement National de Calcul Intensif- A0022A10103).  The authors also acknowledge the access to HPC resources of French regional computing center of Normandy named CRIANN (Centre R\'{e}gional Informatique et d'Applications Num\'{e}riques de Normandie) under the allocations 2017002. The authors gratefully acknowledge financial support from LabEx “Laboratoires d’Excellence” project EMC3 “Energy Materials and Clean Combustion Center” through the project INTRA (LabEx EMC3-2016-PC ). Our work has also benefited from helpful discussions with Prof. A. P. Willis (University of Sheffield, UK), who suggested this particular form for the vortex perturbation.

\end{acknowledgments}

\bibliography{biblioJP}
\bibliographystyle{jfm}
\end{document}